\input{aipcheck}

\documentclass[
    ,final            
  ]
  {aipproc}

\layoutstyle{6x9}

\begin{document}

\title{Radioactivity of the Key Isotope $^{44}$Ti in SN~1987A}

\author{Yuko MOTIZUKI}{
  address={RIKEN, Hirosawa 2-1, Wako 351-0198 Japan}\thanks{E-mail: motizuki@riken.jp.
Spelling of her name (Mochizuki) has changed to Motizuki. }
}

\author{Shiomi KUMAGAI}{
  address={Department of Physics, Faculty of Science and Technology,
Nihon University\\ Kanda-Surugadai 1-8, Chiyoda-ku, Tokyo 101-0062 Japan}
}


\begin{abstract}
We investigate radioactivity from the decay sequence of $\rm ^{44}Ti$
in a young supernova remnant SN~1987A.
We perform Monte-Carlo simulations of degradation of the nuclear lines
to explain a late-time bolometric luminosity which is estimated from 
optical and near-infrared observation at 3600 days after the explosion. 
Assuming the distance to LMC in between 
45.5 and 52.1 kpc,  
we have obtained the initial $\rm ^{44}Ti$ mass of 
$(0.82-2.3) \times 10^{-4} M_\odot$ within the current uncertainty 
of the physical quantities.
The resulting fluxes of $\gamma$- and hard X-rays emerged from the 
$\rm ^{44}Ti$ decay are estimated and compared with the line sensitivity 
of the INTEGRAL/SPI on board and that of NeXT X-ray satellite
planned to be launched in 2010.
The effect of $\rm ^{44}Ti$ ionization on the estimated fluxes
is briefly remarked. 
\end{abstract}

\maketitle


\section{Introduction}

To detect the nuclear $\gamma$-rays from the decay sequence of 
$\rm ^{44}Ti$ is one of the prime target in the current and
future $\gamma$-ray and X-ray satellites.
In particular, SN 1987A, appeared 16 years ago
in the Large Magellanic Cloud (LMC), 
is providing us with a challenge as the one of
such targets.

The important feature of detecting $\rm ^{44}Ti$ nuclear lines 
from young supernova remnants can be summarized as follows: 
The initial yield of $\rm ^{44}Ti$ that is synthesized by a single event of a 
core-collapse supernova explosion is very crucial to constrain 
dynamics of core-collapse supernova nucleosyntehsis.
This is because $\rm ^{44}Ti$ is synthesized at the vicinity of 
the so-called mass cut, that divides the matter which accretes on a 
compact object and the ejecta which is scattered into interstellar space.    
For this, the initial mass of $\rm ^{44}Ti$ depends sensitively on 
1) the location of the mass cut,
2) the maximum temperature and the maximum density behind the shock wave,
and 3) the internal structure 
(\mbox{\raisebox{0.3ex}{$<$}\raisebox{-0.7ex}{\hspace*{-0.8em}$\sim$}}\/
2 $\rm M_{\odot}$ from the center) of a progenitor.

So far, 1.16 MeV nuclear line that is emitted from the decay-chain of $\rm ^{44}Ti$ 
(see below) was detected from Cassiopeia~A 
with COMPTEL/CGRO experiment 
\cite{I94, S00}
and this was confirmed with BeppoSAX
by detecting associated 67.9 and 78.4 keV nuclear lines \cite{V01}.
It should be noted here that RXJ0852-4622, the ``Vela Junior'' remnant,
was first discovered by the 1.16 MeV $\gamma$-ray line as a point source 
\cite{I98}, and then discovered in X-rays
\cite{A98}.   
In the near future, $\rm ^{44}Ti$ nuclear lines are expected to be 
detected from the other young 
(\mbox{\raisebox{0.3ex}{$<$}\raisebox{-0.7ex}{\hspace*{-0.8em}$\sim$}}\/ 
a few$\times$1000 yrs) galactic supernova remnants 
and SN~1987A in LMC.
Further, galactic survey of $\rm ^{44}Ti$ nuclear lines may dig out unknown 
supernova remnants which is difficult to be caught in the other electromagnetic 
wavelengths, and may give us even a scrap of 
information on the galactic supernova rate.

The decay sequence of the radioactive $\rm ^{44}Ti$ is the following:
$\rm ^{44}Ti$ decays by orbital electron capture 
mainly to the second excited state of $\rm ^{44}Sc$ (branching ratio of 99.3\%).
The decay is soon followed by the emissions of 67.9 keV and 78.4 keV nuclear 
deexcitation lines to the ground state of $\rm ^{44}Sc$.
Although until recently the halflife of $\rm ^{44}Ti$ showed a large uncertainty
in those measured in laboratories,
compilation of recent 8 experiments which were performed after
1998 (see., e.g., \cite{H01} and references therein;
\cite{F00})
gives weighted mean halflife of $t_{1/2} = 60 \, \pm$ 1 yr 
(the error is 1 $\sigma$, statistical).
The daughter nucleus, $\rm ^{44}Sc$, then decays almost exclusively 
by positron emission to the first excited state of $\rm ^{44}Ca$,
which emits 1.16 MeV deexcitation line to the ground state. 
The ground state of $\rm ^{44}Ca$ is stable.
The emitted positron in the above sequence ends up with 511 keV 
annihilation line.
It is noted that 
the halflife of $\rm ^{44}Sc$ is merely 3.93 hrs, 
so that the timescale and hence the radioactivity of the whole 
decay chain is regulated by the halflife of $\rm ^{44}Ti$.


We note that laboratory experiments measure the halflife of 
{\em neutral} $\rm ^{44}Ti$.
The crucial point here is that $\rm ^{44}Ti$ decays only
by orbital electron capture. 
This is because the decay Q-value from the ground state of $\rm ^{44}Ti$ 
to the second excited state of $\rm ^{44}Sc$
is less than twice the electron rest mass, which is at least required
for positron emission to be allowed 
by producing two 511 keV $\gamma$-photons when a positron annihilates 
with an electron (and so does that to the first excited state of 
$\rm ^{44}Sc$ for the rest of the minor fraction of the branch).
Actually the halflife of highly ionized $\rm ^{44}Ti$ becomes longer
than that of neutral $\rm ^{44}Ti$ and this affects the radioactivity. 
Thus, we should be careful to apply the experimental 
halflife to this problem:
The electric environment of $\rm ^{44}Ti$ in a young supernova remnant may be
different from that in laboratories.

Previous studies on the $\rm ^{44}Ti$ ionization effect
on its radioactivity are found in references \cite{M99, M01};
these mainly discuss Cassiopeia A. 
A recent paper \cite{MK03} includes a linear analysis in order 
to simply show why and how the $\rm ^{44}Ti$ ionization affects its 
radioactivity.  
As we shall see later, there is a clear possibility of ionization of 
$\rm ^{44}Ti$ ongoing in SN~1987A.
In this article, we first derive the expected nuclear fluxes
without consideration of the $\rm ^{44}Ti$ ionization.
We then briefly discuss how the possibly-ongoing $\rm ^{44}Ti$ ionization 
may change this estimate in the near future, using the result of 
the linear analysis presented in \cite{MK03}.

\section{$\rm \bf ^{44}Ti$ Radioactivity in SN 1987A}

\begin{figure}
  \includegraphics[height=.5\textheight]{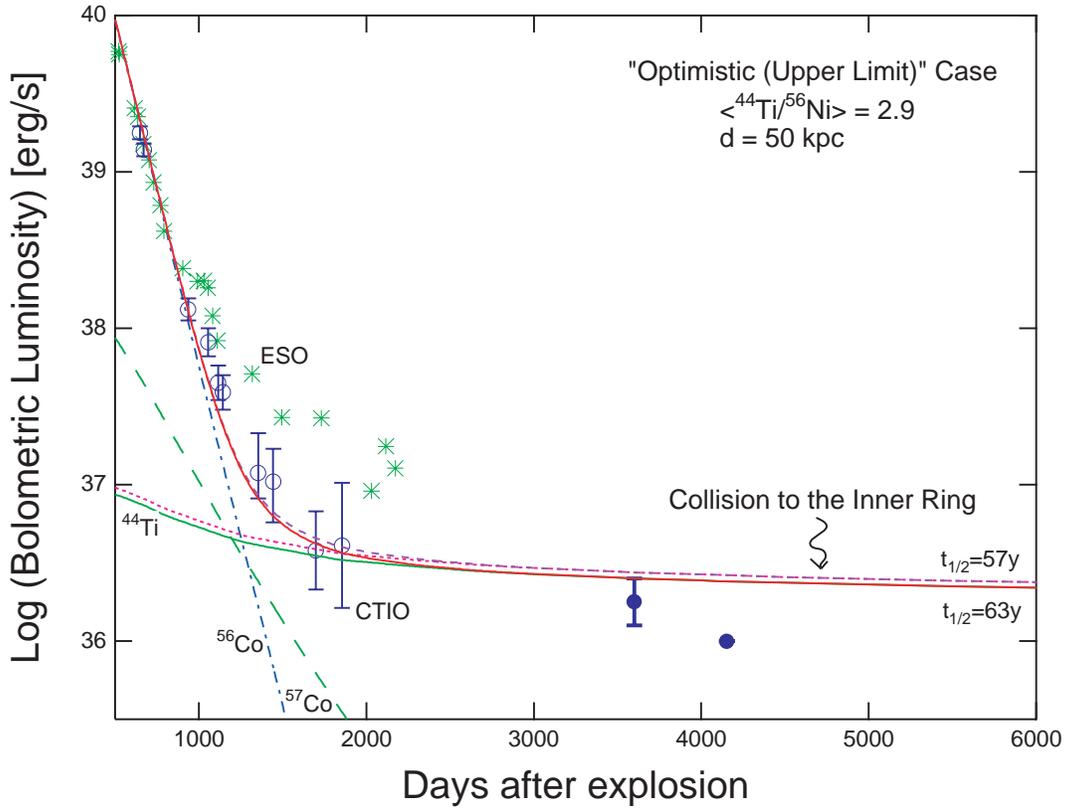}
  \caption{Evolution of the bolometric luminosity of SN~1987A. 
  Observed luminosity data are shown (see the text).
  The thick solid line denotes a theoretical light curve with the $^{44}$Ti halflife
  of $t_{1/2} = 63$ yrs and the short-dashed line denotes that with $t_{1/2} = 
  57$ yrs.  
  The long-dashed line shows a theoretical contribution from the $^{57}$Co decay,
  and the dash-dotted line that from the $^{56}$Co decay.
  The dotted and thin solid lines denote the contributions from the $^{44}$Ti decays that
  correspond to $t_{1/2} = 57$ yrs and 63 yrs, respectively.  
  }
\end{figure}

Figure~1 depicts observation-based data of bolometric luminosity
and theoretically calculated light curves. 
The observed light curve in early
time is first governed by the radioactive decay of 
$^{56}$Ni ($t_{1/2}$ = 6.1 d)
and then that of its daughter $^{56}$Co ($t_{1/2}$ = 77.3 d).  
The synthesized
$^{56}$Co nuclide decays mainly by positron emission to stable $^{56}$Fe.  
Until $\sim$~800 days after the explosion, 
the light curve observed in the
wavelength ranging from ultraviolet to infrared has been
successfully modeled with the energy supply from the decay of
$^{56}$Co 
(see \cite{K93} for details).  

Afterwards, the decline of the observed light curve apparently slowed
down, suggesting the presence of the other heating source.  
As the next source, $^{57}$Co ($t_{1/2}$ = 272 d) plays a role in the
luminosity.  
This epoch does not last long, however.  
The slowness of the decline of the observed light curve becomes
distinguished in particular after $\sim$~1500 days
(see Fig.~1).
The dominant energy source at this moment is believed to be 
the $^{44}$Ti decay.

Suntzeff \cite{S97} reported a bolometric luminosity at 3600 days,
i.e., 10 years after the explosion.
He used the bolometric corrections to the optical colors VK 
to estimate a bolometric magnitude under the assumption that 
the flux distribution was indeed frozen. 
Under this assumption, he obtained a bolometric luminosity,
$
L_{\rm S97} = (1.9 \pm 0.6)  \times 10^{36} \,  {\rm erg \, sec^{-1}}, 
$ 
at 3600 days after the explosion. 
In the following, this bolometric luminosity $L_{\rm S97}$ is referred to as the S97 
bolometric luminosity.  
The author also estimated a bolometric luminosity of 
$\sim~1.0 \times 10^{36} \, {\rm erg \, sec^{-1}}$
at 4151 days after the explosion under the assumption that
the same bolometric corrections from day 1800 \cite{S02}.
The above-mentioned observational data and observation-based values of
bolometric luminosity are shown in Fig.~1.

To explain the upper and the lower bound of the S97 bolometric luminosity 
at 3600 days after the explosion \cite{S97},
we performed Monte-Carlo simulations of Compton degradation of the 
nuclear $\gamma$-photons of
$^{57}$Co (14 keV, 122
keV, 136 keV, etc.) and $^{44}$Ti (68 keV, 78 keV, 511 keV, 1.16 MeV).
The UV, optical, and IR photons originate from the energy loss of the
emitted $\gamma$-rays during the radiative transfer in the ejecta.

To determine the velocity distribution of
particles, we adopt the explosion model 14E1 
\cite{SN90}
whose main-sequence mass, the ejecta mass, and the explosion
energy are 20 $\rm M_{\odot}$, 14.6 $\rm M_{\odot}$ (4.4 $\rm M_{\odot}$ 
core material plus
10.2 $\rm M_{\odot}$ hydrogen-rich envelope), and $1 \times 10^{51}$ erg,
respectively.  
This model was derived from a detailed analysis of the
plateau shape of the light curve of SN~1987A which was observed until
120 days after the explosion, and well accounts for the earlier
optical, X-ray, and $\gamma$-ray light curves of SN 1987A
\cite{N91}. 
Note that 
the $^{56}$Ni mass in SN~1987A has been constrained as $0.07 M_\odot$
from the intensity during the observed exponential decline.

Note also that at the period of the S97 observation the ionization of $\rm ^{44}Ti$
was not relevant.
Later, the supernova blast shock crashed into the dense inner ring,
and the shock heating started to ionize the elements
(see below). 
We therefore use the experimental halflife of neutral $\rm ^{44}Ti$: 
$t_{1/2} = 60 \pm 3$ yrs (3$\sigma$ deviation)
to explain the S97 bolometric luminosity.
The distance to SN~1987A is adopted to be $48.8 \pm 3.3$ kpc 
(3$\sigma$, see \cite{GU98}).
Other details of our calculation are found in \cite{K93}; 
in the present study, the adopted nuclear decay parameters 
have been updated.

In Fig.~1, two calculated bolometric light curves which stick to the 
upper bound of the S97 luminosity are also shown.
The difference of the two theoretical curves is in the adopted halflife of $\rm ^{44}Ti$;
the one is calculated with $t_{1/2} = 57$ yrs and the other with $t_{1/2} = 63$ yrs. 
One sees in Fig.~1 that the current uncertainty in the $\rm ^{44}Ti$ 
halflife no longer produces a remarkable difference in the light curve.

\begin{figure}
  \includegraphics[height=.5\textheight]{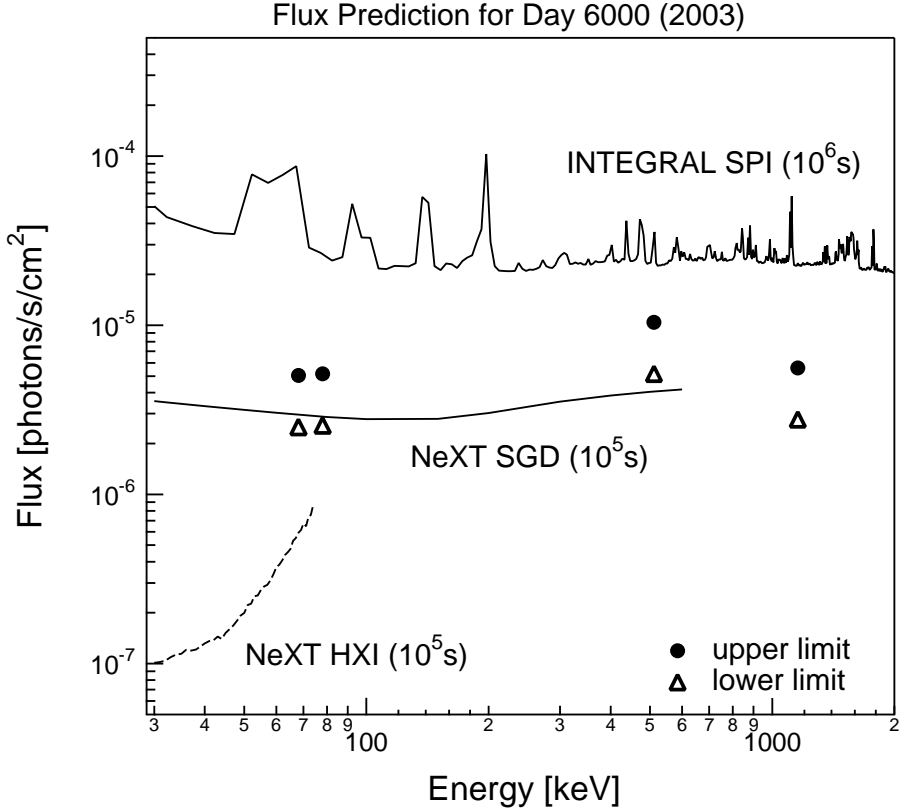}
  \caption{Prediction of the nuclear fluxes associated with the $\rm ^{44}Ti$ decay
in SN~1987A at 6000 days after the explosion (i.e. in 2003). 
The line sensitivity of INTEGRAL SPI on board for the typical observation time
$10^6$ sec and that of the NeXT mission for the typical observation time of $10^5$ sec
are also shown for comparison.  
The NeXT satellite is planned to be launched in 2010.
}
\end{figure}

The upper-bound light curve gives 
$<^{44}$Ti/$^{56}$Ni$> = 2.9$,
where $<^{44}$Ti/$^{56}$Ni$>$ is defined as 
$<^{44}$Ti$/^{56}$Ni$> \equiv [X(^{44}{\rm
Ti}) / X(^{56}{\rm Ni})]_{\rm 87A} / [X(^{44}{\rm Ca}) / X(^{56}{\rm
Fe})]_\odot$,
i.e.,
the ratio of $^{44}$Ti/$^{56}$Ni
in amount in SN~1987A to $^{44}$Ca/$^{56}$Fe in the
solar neighborhood.
The obtained value 2.9 for the upper bound is considered to be acceptable. 
One of the reasons for this is that 
the synthesized amount of $^{44}$Ti theoretically tends to be more 
abundant in aspherical supernovae than in spherical ones;
actually recent Hubble Space Telescope images and spectroscopy 
have revealed that SN~1987A has an aspherical geometry \cite{W02}.
Translated from the obtained $<^{44}$Ti/$^{56}${\rm Ni}> values,
we find the initial $\rm ^{44}Ti$ mass of
$(0.82-2.3) \times 10^{-4} M_\odot$ 
within the known uncertainty of the experimental values mentioned above.

Figure~2 shows the expected nuclear fluxes at 6000 days after the explosion
(i.e., in 2003) for both the obtained upper 
and the lower bounds of the initial $\rm ^{44}Ti$ mass. 
It is worthwhile to mention here that 
the obtained $\rm ^{44}Ti$ masses depend on the assumed 
distance to SN~1987A, but the expected fluxes do not.
In Fig.~2, the line sensitivity of INTEGRAL/SPI on board and 
that of the planned NeXT mission are also shown for comparison.

We see in Fig.~2 that, with $10^6$ sec observation time, it might be difficult
for INTEGRAL/SPI to detect the $\rm ^{44}Ti$ nuclear lines from SN~1987A.
We note that our upper bound estimate relies on
the upper bound of the S97 bolometric luminosity, which was estimated
under the assumption described previously and corresponds to usual 
1 $\sigma$ deviation: There is a possibility that the starting point, 
the bolometric luminosity itself, may be larger. 
Theoretically, on the other hand, the larger bolometric luminosity
corresponds to the larger $<^{44}$Ti/$^{56}$Ni$>$ value.
The feasibility of the largest $<^{44}$Ti/$^{56}$Ni$>$ value 
remains to be solved, but too large value could not be naturally adopted.

Finally, we point out that the ionization process of $\rm ^{44}Ti$ is
considered to be well underway in SN 1987A, due to
shock heating caused by the collision of the supernova blast shock with 
the dense inner ring:
H-like and He-like ionization stages of O, Ne, Mg, and 
Si have been already observed, and also SN~1987A is observed to be
a very rapidly evolving remnant 
(see \cite{B00, M02}).
If $\rm ^{44}Ti$ reaches the high-ionization stages in the future, 
the expected fluxes shown in Fig.~2 will become smaller, 
as discussed in the linear analysis presented in \cite{MK03}.
To get a rough idea, if {\em all} the $\rm ^{44}Ti$ should be 
in the He-like (H-like) ionization state, 
the radioactivity of SN~1987A would suffer a $\sim$ 9 (45) \% reduction
(see Table~1 of \cite{MK03}).    
Under the realistic situation in which the ionization of the elements 
evolves by the shock heating, various ionization stages will be led.
Even in such a case, 68, 78, 511 keV lines appear to be detectable
with a planned detector on a future mission such as NeXT (see Fig.~2).
Actual conclusion of the effect of the ionization on the radioactivity
requires the knowledge of the temperature and the density evolution 
of the supernova remnant, and this will be a future subject for 
SN~1987A.


\begin{theacknowledgments}
We are grateful to 
J. Kn\"{o}dlseder for providing us with the SPI sensitivity
and
T. Takahashi
also for providing us with the planned line sensitivity of 
the NeXT mission.
Y.M. would like to thank
P. Leleux and 
A. Gould
for helpful comments.
\end{theacknowledgments}


\bibliographystyle{aipproc}   

\bibliography{sample}

\IfFileExists{\jobname.bbl}{}
 {\typeout{}
  \typeout{******************************************}
  \typeout{** Please run "bibtex \jobname" to optain}
  \typeout{** the bibliography and then re-run LaTeX}
  \typeout{** twice to fix the references!}
  \typeout{******************************************}
  \typeout{}
 }

\end{document}